\documentclass[aps,prl,reprint,superscriptaddress,nobibnotes]{revtex4-2}

\usepackage{graphicx}
\usepackage{csquotes}
\usepackage{xcolor} 
\usepackage{amsmath}

\renewcommand{\paragraph}[1]{\textit{#1}. \nobreak\ignorespaces}

\begin{document}

\graphicspath{ {./figures/} }

\title{ Slice Emittance Preservation and Focus Control in a Passive Plasma Lens }

\author{J. Bj\"{o}rklund Svensson}
\email{jonas.bjorklund\_svensson@fysik.lu.se}
\affiliation{Deutsches Elektronen-Synchrotron DESY, Notkestra{\ss}e 85, 22607 Hamburg, Germany}
\affiliation{Department of Physics, Lund University, P.O. Box 118, SE-22100 Lund, Sweden}%

\author{J. Beinortait\.{e}}
\affiliation{Deutsches Elektronen-Synchrotron DESY, Notkestra{\ss}e 85, 22607 Hamburg, Germany}%
\affiliation{University College London, Gower St., London WC1E 6BT, United Kingdom}%

\author{L. Boulton}
\affiliation{Deutsches Elektronen-Synchrotron DESY, Notkestra{\ss}e 85, 22607 Hamburg, Germany}%

\author{B. Foster}
\affiliation{Deutsches Elektronen-Synchrotron DESY, Notkestra{\ss}e 85, 22607 Hamburg, Germany}%
\affiliation{Universit\"{a}t Hamburg, Luruper Chaussee 149, 22761 Hamburg, Germany}%
\affiliation{John Adams Institute, Department of Physics, University of Oxford, Oxford, UK}

\author{J. M. Garland}
\affiliation{Deutsches Elektronen-Synchrotron DESY, Notkestra{\ss}e 85, 22607 Hamburg, Germany}%

\author{P. Gonz\'{a}lez Caminal}
\affiliation{Deutsches Elektronen-Synchrotron DESY, Notkestra{\ss}e 85, 22607 Hamburg, Germany}%

\author{M. Huck}
\affiliation{Deutsches Elektronen-Synchrotron DESY, Notkestra{\ss}e 85, 22607 Hamburg, Germany}%

\author{H. Jones}
\affiliation{Deutsches Elektronen-Synchrotron DESY, Notkestra{\ss}e 85, 22607 Hamburg, Germany}%

\author{A. Kanekar}
\affiliation{Deutsches Elektronen-Synchrotron DESY, Notkestra{\ss}e 85, 22607 Hamburg, Germany}%

\author{G. Loisch}
\affiliation{Deutsches Elektronen-Synchrotron DESY, Notkestra{\ss}e 85, 22607 Hamburg, Germany}%

\author{J. Osterhoff}
\email{Present address: Lawrence Berkeley National Laboratory, Berkeley, CA 94720, USA}
\affiliation{Deutsches Elektronen-Synchrotron DESY, Notkestra{\ss}e 85, 22607 Hamburg, Germany}%

\author{F. Pe\~{n}a}
\email{Present address: Ludwig-Maximilians–Universität München, Am Coulombwall 1, 85748 Garching, Germany}
\affiliation{Deutsches Elektronen-Synchrotron DESY, Notkestra{\ss}e 85, 22607 Hamburg, Germany}%
\affiliation{Universit\"{a}t Hamburg, Luruper Chaussee 149, 22761 Hamburg, Germany}%

\author{S. Schr\"{o}der}
\email{Present address: Lawrence Berkeley National Laboratory, Berkeley, CA 94720, USA}
\affiliation{Deutsches Elektronen-Synchrotron DESY, Notkestra{\ss}e 85, 22607 Hamburg, Germany}%

\author{M. Th\'{e}venet}
\affiliation{Deutsches Elektronen-Synchrotron DESY, Notkestra{\ss}e 85, 22607 Hamburg, Germany}%

\author{S. Wesch}
\affiliation{Deutsches Elektronen-Synchrotron DESY, Notkestra{\ss}e 85, 22607 Hamburg, Germany}%

\author{M. Wing}
\affiliation{Deutsches Elektronen-Synchrotron DESY, Notkestra{\ss}e 85, 22607 Hamburg, Germany}%
\affiliation{University College London, Gower St., London WC1E 6BT, United Kingdom}%

\author{J. C. Wood}
\email{jonathan.wood@desy.de}
\affiliation{Deutsches Elektronen-Synchrotron DESY, Notkestra{\ss}e 85, 22607 Hamburg, Germany}%

\author{R. D'Arcy}
\email{richard.darcy@physics.ox.ac.uk}
\affiliation{Deutsches Elektronen-Synchrotron DESY, Notkestra{\ss}e 85, 22607 Hamburg, Germany}%
\affiliation{John Adams Institute, Department of Physics, University of Oxford, Oxford, UK}

\date{July 15, 2026}

\begin{abstract}
Strong, symmetrically focusing plasma lenses are promising for accommodating the small beams associated with plasma-based accelerators and collider final foci. However, while focusing with active and passive plasma lenses has been experimentally demonstrated, compatibility with high-brightness beams relevant for applications has not. In this work, we show experimentally that passive plasma lenses can preserve free-electron-laser-quality slice emittance while focusing two orders of magnitude more strongly than quadrupole magnets, and that the focal parameters can be controlled.
\end{abstract}

\maketitle


\paragraph{Introduction}
In plasma-based accelerators (PBAs) \cite{tajima_1979_PRL,chen_1985_PRL,ruth_1985_PartAcc}, an intense electron bunch or laser pulse (the \enquote{driver}) expels the free plasma electrons from its propagation axis, exciting a plasma wake within which strong electromagnetic fields enable high-gradient acceleration (up to hundreds of GV/m) and focusing (kT/m\,--\,MT/m) of a trailing electron bunch (the \enquote{witness}). These fields greatly exceed those in radio-frequency (RF) accelerators, where electric breakdown in the metallic cavities limits the accelerating gradients to a few hundred MV/m \cite{simakov_2018_NIMA}, and magnet technology limits the focusing gradients to a few hundred T/m \cite{ghaith_2019_instruments}. Plasmas thus offer a possible route to smaller, cheaper accelerator systems through both the accelerating and focusing gradients.

Over the past decades, PBAs have demonstrated major improvements in energy gain \cite{hogan_2005_PRL,blumenfeld_2007_Nature,aniculaesei_2023_MRE,picksley_2024_PRL}, energy efficiency \cite{litos_2014_Nature,lindstrom_2021_PRL,kirchen_2021_PRL,pena_2024_PRR}, energy spread \cite{wang_2016_PRL,lindstrom_2021_PRL,kirchen_2021_PRL}, and acceleration stability and reproducibility \cite{maier_2020_PRX,lindstrom_2021_PRL,wood_2024_ipac,winkler_2025_Nature}. However, while challenging applications such as PBA-driven free-electron lasers (FELs) have been demonstrated for longer wavelengths (IR\,--\,XUV) \cite{wang_2021_Nature,pompili_2022_Nature,labat_2023_NatPhoton,barber_2025_PRL}, the performance of RF-based FEL facilities has not yet been matched; X-ray FELs and linear colliders require a smaller normalized beam emittance ($\lesssim1$\,mm$\cdot$mrad, see Appendix). 

In a blown-out plasma wake, the focusing gradients are linear, and emittance preservation is possible \cite{rosenzweig_1991_PRA,lu_2006_PRL,lindstrom_2024_NatCommun}. The natural beam size in this regime is given by the matched beta function \cite{courant_1958_annphys}, $\beta_\mathrm{m} = \sqrt{ 2 \epsilon_0 E_\mathrm{b} / (n_\mathrm{p}e^2) }$, where $E_\mathrm{b}$ is the beam energy, $e$ is the unit charge, $n_\mathrm{p}$ is the plasma electron density, and $\epsilon_0$ is the vacuum permittivity; this is typically on the sub-mm to cm scale for GeV-class PBA beams, compared to tens of meters in RF accelerators. The strong focusing required to accommodate these small beams can cause deleterious chromatic aberration, yielding a relative emittance growth of $\Delta\varepsilon/\varepsilon_0 \approx L^{*2}\sigma_\delta^2/(2\beta_\mathrm{m}^2)$ (see Appendix), where $L^*$ is the distance between the beam waist and the optic and $\sigma_\delta$ is the root-mean-square (RMS) relative energy spread. A small beam size usually corresponds to a large divergence, which drives emittance growth even in free-space propagation \cite{migliorati_2013_PRSTAB}. Therefore, minimizing the waist-optic distance is advantageous, a distance limited by the linear focusing strength, $\kappa = 1/(lL^*)$, where $l$ is the length of the optic. High strength simultaneously enables more compact optical systems, which currently drive space requirements of PBA experiments. Plasma lenses could realize these advantages. 

Compared to quadrupole magnets, plasma lenses provide higher focusing gradients and symmetric focusing. Active plasma lenses (APLs) focus using the magnetic field induced by a discharge current \cite{vanTilborg_2015_PRL}, while passive plasma lenses (PPLs), which are similar to PBAs, use the transverse plasma wakefield \cite{chen_1987_PartAcc}. Emittance preservation and kT/m focusing gradients have been demonstrated in APLs \cite{lindstrom_2018_PRL,vanTilborg_2015_PRL,sjobak_2021_PRAB}, but their compatibility with high-brightness beams is limited by wake excitation and Coulomb scattering \cite{vanTilborg_2018_PoP,lindstrom_2018_arxiv,pompili_2024_PRE,lindstrom_2022_jinst}. In contrast, because they rely on plasma wakes and can be placed nearer the focus, PPLs can be operated with lower-atomic-number gases and smaller beam sizes, reducing scattering-driven emittance growth. Their higher focusing gradients (up to MT/m \cite{ng_2001_PRL}) allow for placement closer to the focus. These attributes suggest that PPLs could act as an ideal bridge between magnetic optics and PBAs, potentially facilitating plasma-accelerator staging \cite{lindstrom_2021_PRAB}. However, while passive plasma lensing has been previously demonstrated \cite{ng_2001_PRL,thompson_2010_PoP,thaury_2015_NatComms,kuschel_2016_PRAB,marocchino_2017_APL,bjorklund_svensson_2021_NatPhys,chang_2023_PRA,gustafsson_2024_SciRep}, beam-quality preservation has not.

In this work, we experimentally demonstrate slice-emittance preservation at sub-mm$\cdot$mrad emittances and focus control at cm-scale waist betas with an electron-beam-driven PPL, using beams with three orders of magnitude higher brightness than earlier work \cite{lindstrom_2018_PRL}. This was achieved by operating with a highly underdense plasma ($10^{14}$\,cm\textsuperscript{-3}), which decreased the demands on the driver to reach a full blowout and limited the change in particle energy to the per-mille level.


\begin{figure}[]
\includegraphics[width=1.0\linewidth]{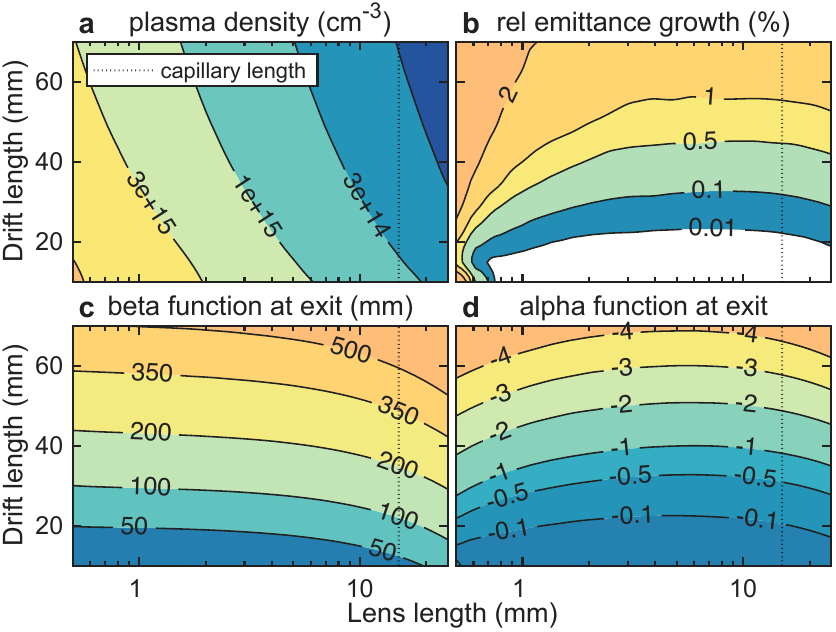}
\caption{\label{fig:1} Plasma and beam parameters from the preparatory PIC simulations with idealized beams and plasma. \textbf{a}. Plasma densities. \textbf{b}. Relative emittance growth. \textbf{c}. Final beta functions. \textbf{d}. Final alpha functions. The dashed line marks the 15-mm lens length used in the experiment.}
\end{figure}

\paragraph{Limiting the parameter space}
A set of preparatory particle-in-cell (PIC) simulations were performed (see Appendix) to find the limits of the lens operating range, where the driver could no longer sustain an emittance-preserving wake. Because the nominal beam waist parameters were considered known \cite{lindstrom_2024_NatCommun}, the reciprocal of the intended experiment, collimation of a beam diverging from this beam waist, was simulated to simplify the procedure and interpretation. A grid of lens lengths, $l$, and distances from waist to lens (\enquote{drift lengths}), $L^*$, was generated. At each grid point, the collimating focusing strength was found by numerically solving 
\begin{equation*}
    \kappa \left[\beta^{*2}+L^{*2}\right] + \sqrt{\kappa}\,L^*\left[\tan{(\sqrt{\kappa}\,l)} - \cot{(\sqrt{\kappa}\,l)}\right] = 1
\end{equation*}
for $\kappa$, obtained from thick-lens matrix theory by requiring $\alpha=0$ after the lens \cite{lee_2019_AcceleratorPhysics}, resulting in a collimated beam. The corresponding plasma densities, $n_\mathrm{p} = 2\epsilon_0E_\mathrm{b}\kappa/e^2$, were in the range $5\times10^{13}$\,--\,$1\times10^{16}$\,cm\textsuperscript{-3}, see Fig. \ref{fig:1}a.

Figures \ref{fig:1}b, c, and d show the relative emittance growth between the start and end of the lens, and the final beta and alpha functions, respectively. The presented values are the longitudinal slice averages, to suppress chromatic effects from any energy spread induced by the lens, highlighting wake excitation limitations. These limitations are evident in Figs. \ref{fig:1}b and d, showing varying degrees of emittance growth and suboptimal collimation ($\alpha<0$) from insufficient wake excitation. The optimal lens length was found to be around 6\,mm, within a wide range of approximately 1-20\,mm. A 15-mm lens capillary, previously manufactured and tested for a different experiment, was therefore selected for use in experiments.

\begin{figure}[b]
\includegraphics[width=1.0\linewidth]{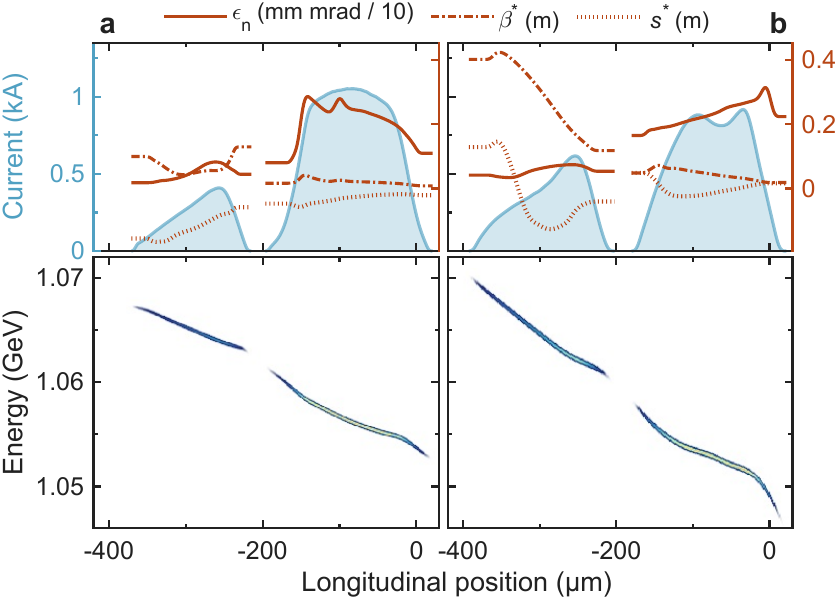}
\caption{\label{fig:2} Longitudinal phase spaces and horizontal slice parameters before plasma interaction, reconstructed from experimental data (see Appendix). Beam used with: \textbf{a}. nitrogen (Figs. \ref{fig:3} and \ref{fig:4}), \textbf{b}. argon (Figs. \ref{fig:5} and \ref{fig:6}). }
\end{figure}

\paragraph{Experimental setup}
The experiments were performed at the FLASHForward beamline at DESY using electron bunches from the FLASH linear accelerator (linac) \cite{darcy_2019_RSTA,schreiber_2015_HPLSE}. The 900-pC bunches were accelerated to 1\,GeV in superconducting RF cavities and compressed to 1.3\,ps full-width (1\,kA peak current) by two magnetic chicanes. The driver-witness bunch pairs were generated via a collimator in a dispersive section \cite{schroder_2020_JPhysConf}; this was enabled by the linear longitudinal (time-energy) phase space (LPS) achieved with a harmonic cavity in the linac (see Fig. \ref{fig:2}).

\begin{figure}[tb]
\includegraphics[width=1.0\linewidth]{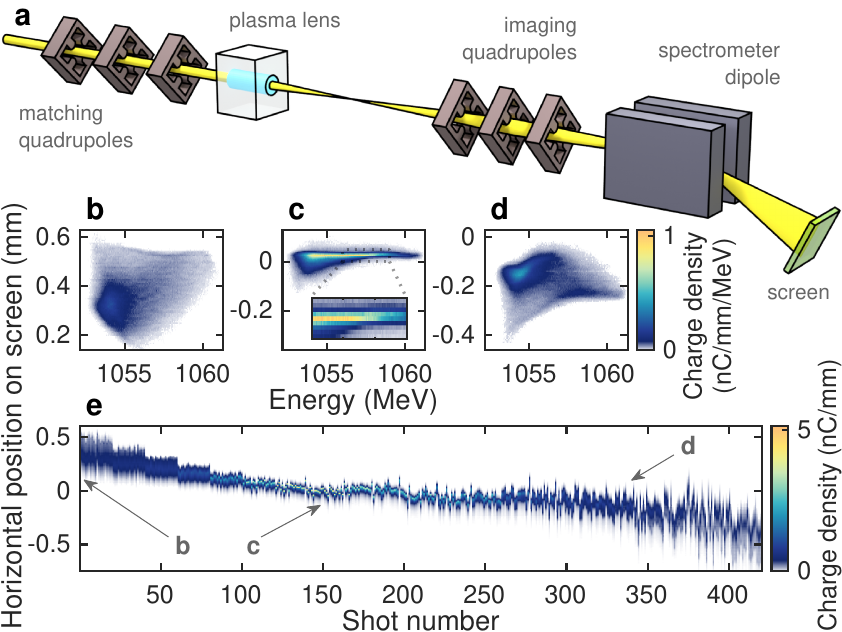}
\caption{\label{fig:3} Experimental setup and raw data examples. \textbf{a}. The core of the experimental setup. The two bunches, propagating left to right, are pre-focused into the PPL, which is then used to focus further. The bunches are then imaged onto the spectrometer screen. \textbf{b}-\textbf{d}. Spectrometer images at three different quadrupole imaging settings along a measurement scan, with the horizontal projections of all images shown in \textbf{e}. }
\end{figure}

Figure \ref{fig:3}a shows the core of the experimental setup. A set of quadrupole magnets upstream of the interaction chamber was used to match the beam into the plasma lens with a waist beta, $\beta^*$, of approximately 100\,mm. In plasma-acceleration-type experiments, the focus is typically significantly tighter with a waist beta of 10\,--\,20\,mm \cite{lindstrom_2024_NatCommun}. The lens was then used to focus the beam further, as if into a hypothetical downstream PBA stage.

The plasma was generated by high-voltage (HV; 20\,kV, 500\,ns FWHM, 490\,A) discharges through a nitrogen-filled (1\,mbar), 15-mm-long, 1.5-mm-diameter sapphire capillary. Nitrogen was chosen over argon as it has a much lower atomic number, reducing Coulomb scattering \cite{lindstrom_2022_jinst}, while being similarly easy to ionize. It was chosen in preference to a lighter gas, e.g. hydrogen, because of the smaller ion motion, which is also suppressed by the larger beam size compared to that used inside the PBA \cite{rosenzweig_2005_PRL,an_2017_PRL}. A 15-mm lens yielded an expected plasma density of 1\,--\,$3\times10^{14}$\,cm\textsuperscript{-3}, below the resolution of standard diagnostics \cite{garland_2021_RevSciInstr} (see Appendix).

\begin{figure*}[]
\includegraphics[width=\textwidth]{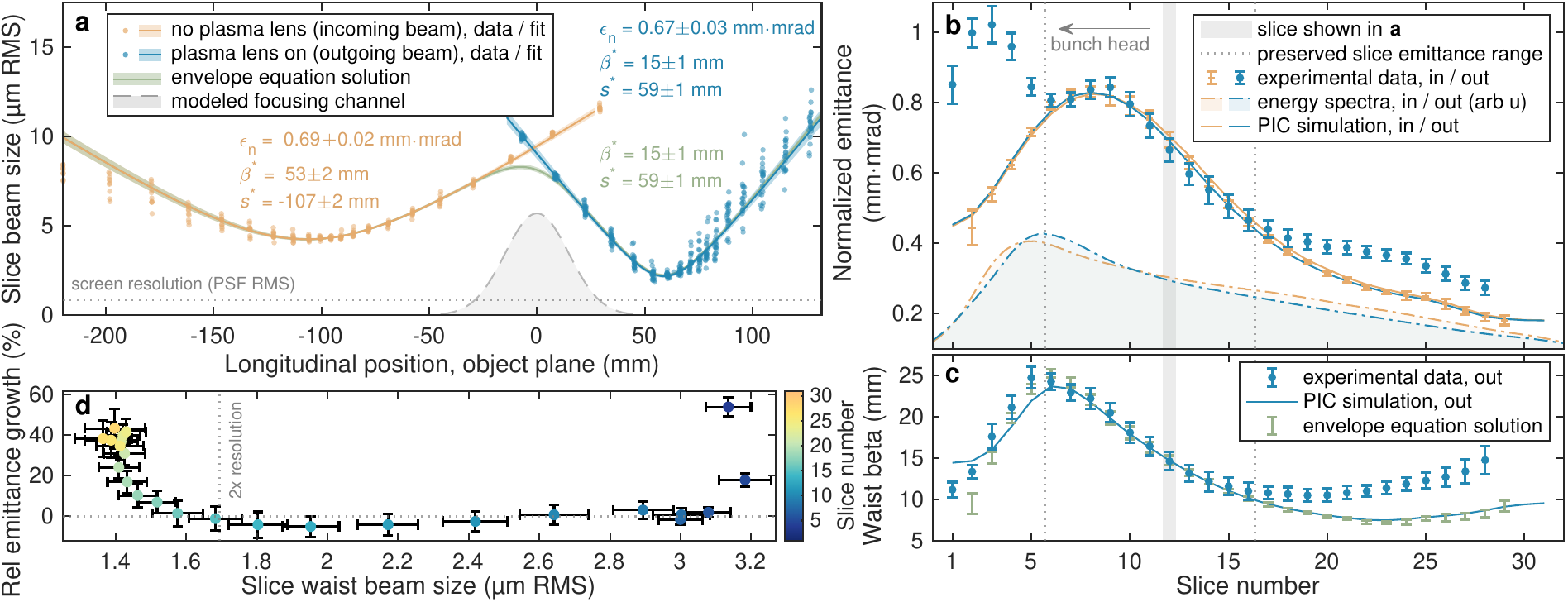}
\caption{\label{fig:4} Slice emittance preservation in nitrogen plasma. \textbf{a}. Measured RMS beam sizes (dots) and corresponding fits (colored lines) for the incoming (orange) and outgoing (blue) witness bunch, along with the corresponding fit parameters, for energy slice 12. The green line shows the median of Monte-Carlo simulations solving Eq. \eqref{eq:betatron} through the Gaussian focusing channel (gray shaded area). Errors show the 95\,\% confidence interval (CI) of the fits. \textbf{b}. Incoming and outgoing slice normalized emittances. \textbf{c}. Outgoing waist betas. The slice emittances are preserved within the dotted lines, within which the measured, modeled, and simulated waist parameters agree. \textbf{d}. Measured relative emittance growth (c.f. \textbf{b},\textbf{c}) plotted against measured waist beam sizes.}
\end{figure*}

\paragraph{Slice emittance preservation}
Measurements of the horizontal distribution of the bunch were performed on the high-resolution imaging electron spectrometer (see Fig. \ref{fig:3}a and Appendix). A type of multi-quadrupole scan, an \enquote{object-plane scan} \cite{lindstrom_2024_NatCommun}, was used to measure the beam parameters. In such a scan, the quadrupole magnets point-to-point image a range of longitudinal positions, $s$, along the beamline onto the spectrometer screen, providing measurements of the transverse RMS beam size, $\sigma(s)$. Figures \ref{fig:3}b\,--\,d and e show example spectrometer images and resulting horizontal beam distributions from such a scan, respectively. Around a beam waist, 
\begin{equation}\label{eq:beamprop}
    \sigma(s) = \sqrt{\frac{\varepsilon_\mathrm{n}\beta^*}{\gamma_\mathrm{b}}\left[ 1 + \left(\frac{s-s^*}{\beta^{*}}\right)^2 \right]}\,\,,
\end{equation}
which was fitted to the measurements to determine the waist beta, the waist location, $s^*$, and the normalized emittance, $\varepsilon_n$; $\gamma_\mathrm{b}$ is the Lorentz factor of the beam. Figure \ref{fig:4}a shows RMS beam sizes and fits for the incoming and outgoing beams in orange and blue, respectively.

Since the object-plane scans were performed on a spectrometer, energy slice information could be obtained. A compromise between energy resolution and signal-to-noise was found with 31 energy slices of equal width (approximately 0.25\,MeV). The transverse distribution of each energy slice and that of the whole beam could be treated similarly, and energy-slice beam parameters could thus be obtained; these are shown in Fig. \ref{fig:4}a for slice 12. The small relative changes in mean energy and RMS energy spread induced by the lens (0.1\,--\,0.2\,\% and 2\,\%, respectively) enabled a direct comparison between incoming and outgoing beams. Figure \ref{fig:4}b shows incoming and outgoing slice emittances and energy spectra, demonstrating horizontal normalized slice-emittance preservation for slices 6\,--\,16, containing half of the total bunch charge.

The small changes in energy also allowed the beam propagation to be modeled using the betatron envelope equation \cite{lee_2019_AcceleratorPhysics}, 
\begin{equation}\label{eq:betatron}
    \frac{1}{2}\beta^{\prime\prime} + \kappa\beta - \frac{1}{\beta}\left[1 + \left(\frac{\beta^\prime}{2}\right)^2 \right] = 0,
\end{equation}
where primes denote differentiation with respect to $s$. Using the covariances from the fits to the incoming beam, 1000 random sets of initial parameters (normalized emittance, waist beta, and waist location) were generated from a tri-variate normal distribution for each slice and propagated through the assumed focusing channel to model the focused waist parameters. The outgoing slice-waist parameters of the emittance-preserved slices could be reproduced (see Fig. \ref{fig:4}a and c) by assuming the action of the PPL to be approximated by a Gaussian focusing channel \cite{schroder_2020_NatCommun,garland_2021_RevSciInstr,lindstrom_2024_NatCommun} with an RMS length of 14.8\,mm and peak focusing strength of $\kappa=780$\,m\textsuperscript{-2}, corresponding to a peak focusing gradient of 2.7\,kT/m and plasma density of $9.1\times10^{13}$\,cm\textsuperscript{-3}. The resulting RMS envelope propagation for the slice indicated in Fig. \ref{fig:4}b is shown in green in Fig. \ref{fig:4}a. PIC simulations using a nearly identical plasma channel and bunch parameters reconstructed from experimental data, where energy slices were mapped to time via the linear LPS, agree with the model and show emittance preservation for the whole bunch (see Fig. \ref{fig:4}b and c, and Appendix).

\paragraph{Reaching the resolution limit}
The energy-slice emittances at the head (left) and tail (right) of Fig. \ref{fig:4}b did not appear to be preserved. The source of the tail emittance growth is alluded to in Fig. \ref{fig:4}c, where the modeled waist betas in the tail are significantly smaller than measured. The variation in downstream waist parameters resulted from varying slice Twiss parameters in the incoming beam (see Fig. \ref{fig:2}), which persisted throughout the experimental campaign. Assuming preserved slice emittances and accurately modeled waist betas, the resulting slice waist beam sizes of the tail are between 0.79 and 1.3\,$\mu$m RMS---near, and even below, the measured resolution of 0.85\,$\mu$m RMS (screen equivalent at the object plane). This limitation is also visible in the inset in Fig. \ref{fig:3}c. Figure \ref{fig:4}d further illustrates the resolution limitation: apart from the head slices, the slice emittances are preserved for beam sizes down to approximately twice the resolution.

The head emittance growth resulted from a small but measurable \enquote{beam tilt} (longitudinal-transverse position correlation, see Figs. \ref{fig:3}b\,--\,d) in the incoming bunch, which increased during focusing, leading to a slight overlap between adjacent slices when projected onto the energy axis. The slice variations were also present upstream at the collimator location, resulting in a suboptimal collimation and a considerable change in waist parameters in the head of the bunch (see Fig. \ref{fig:2}). Coupled with increased tilt after focusing, the increased slice overlap caused slice-emittance growth. It thus seems plausible that the emittance increase seen in the tail and head is a result of the diagnostics and suboptimal beam preparation, respectively, rather than the PPL itself.

\paragraph{Varying the lens parameters}
While the working point in Fig. \ref{fig:4} shows that slice-emittance preservation is possible in PPLs, there is no information on how the PPL behavior scales with plasma parameters, which is important for beam-optics matching. Unfortunately, the sapphire capillary degraded over time, preventing stable discharging in nitrogen. The lens could not be exchanged during the experimental run, but using argon at higher pressures and discharge voltages stabilized the plasma generation enough to continue experimentation.

\begin{figure}[tb]
\includegraphics[width=1.0\linewidth]{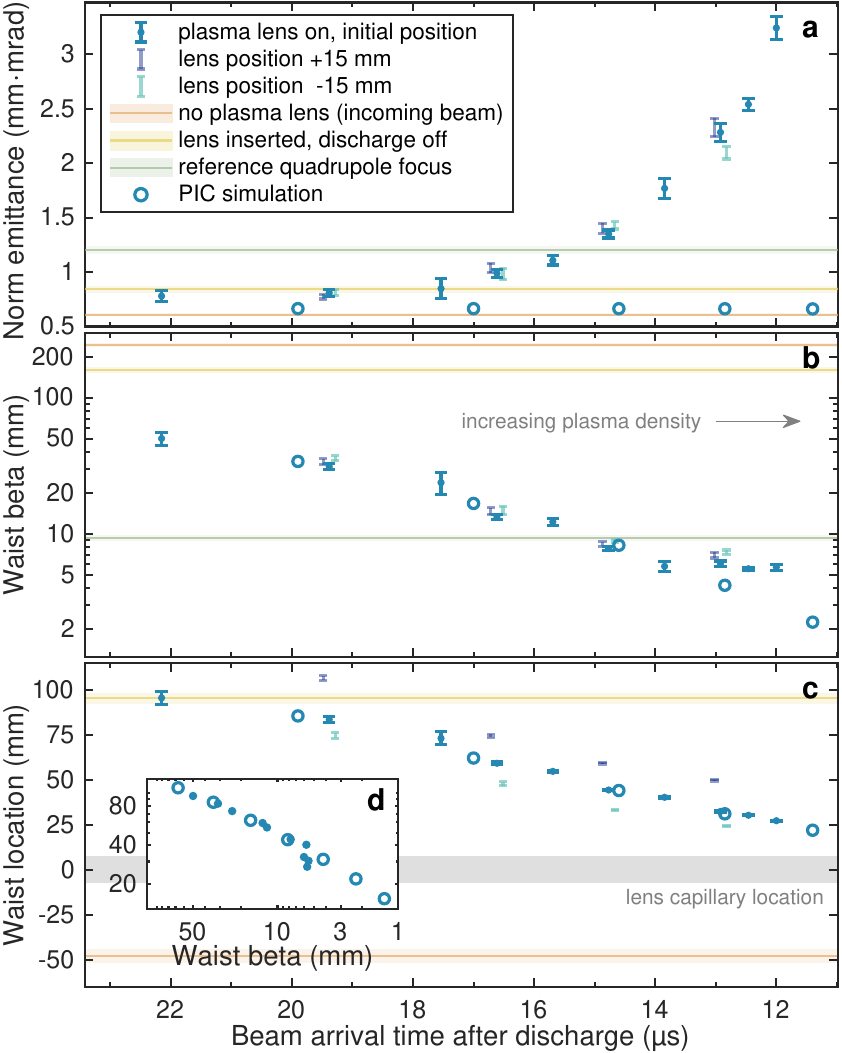}
\caption{\label{fig:5} Projected witness-bunch waist parameters as a function of delay between the discharge in argon and beam arrival, thereby changing the plasma density (see Appendix). \textbf{a}. Normalized emittances. \textbf{b}. Waist betas, \textbf{c}. Waist locations. The reference quadrupole focus is a tight focus using the matching quadrupoles, without the PPL. The horizontal displacement between grouped error bars is for ease of visualization only. Errors show the 95\,\% CI of the fit parameters. \textbf{d}. Measured and simulated waist locations versus waist betas. Error bars removed for clarity. } 
\end{figure}

However, this change also introduced unwanted effects: Figure \ref{fig:5} shows the \enquote{projected} (full-beam) witness waist parameters as a function of the beam arrival time (and thus plasma density, see Appendix), where neutral-gas Coulomb scattering increased the emittance of the beam propagating through the capillary with the discharge off, see Fig. \ref{fig:5}a. Suppressing Coulomb scattering was a main reason for initially operating with nitrogen. Nevertheless, the three lowest plasma densities showed no additional emittance growth, despite significant focusing. Higher plasma densities (stronger focusing) yielded smaller beam waists located closer to the lens, see Fig. \ref{fig:5}b\,--\,d, as intuitively expected. A corresponding decrease in mean energy of 0\,--\,2.5\,MeV$\pm5$\,keV was observed, with only 1.2\,MeV at 14.8\,$\mu$s. The primary cause of observed emittance growth is once again related to resolution effects. Corroboration can be found in the agreement between the PPL data and the reference quadrupole focus, showing similar emittances at equal waist beta. No emittance growth was observed in PIC simulations replicating the waist behavior. Coulomb scattering, affecting individual slices differently because of the incoming slice Twiss variations \cite{muller_2001_note}, prevented a direct slice comparison.

The smallest measured projected (slice) waist beta was 5.6$\pm$0.1\,mm (1.6$\pm$0.1\,mm), a demagnification factor of the incoming waist beta of 44 (166), suggesting a minimum slice RMS beam size below 0.5\,$\mu$m. A larger imaging magnification of around -10 (maximum for the actual object-plane range) enabled these smaller waists to be measured. Moving the lens upstream and downstream by 15\,mm primarily affected the downstream waist location, leaving the emittances and waist betas nearly unaffected. This resulted from the lens being positioned near the large incoming beam waist; the incoming beam optics varied little between different lens positions. The achieved waist betas covered a range relevant for matching into a GeV-class PBA, which, together with the flexible waist positioning, could offer a way of optimizing optics matching.

\begin{figure}[tb]
\includegraphics[width=1.0\linewidth]{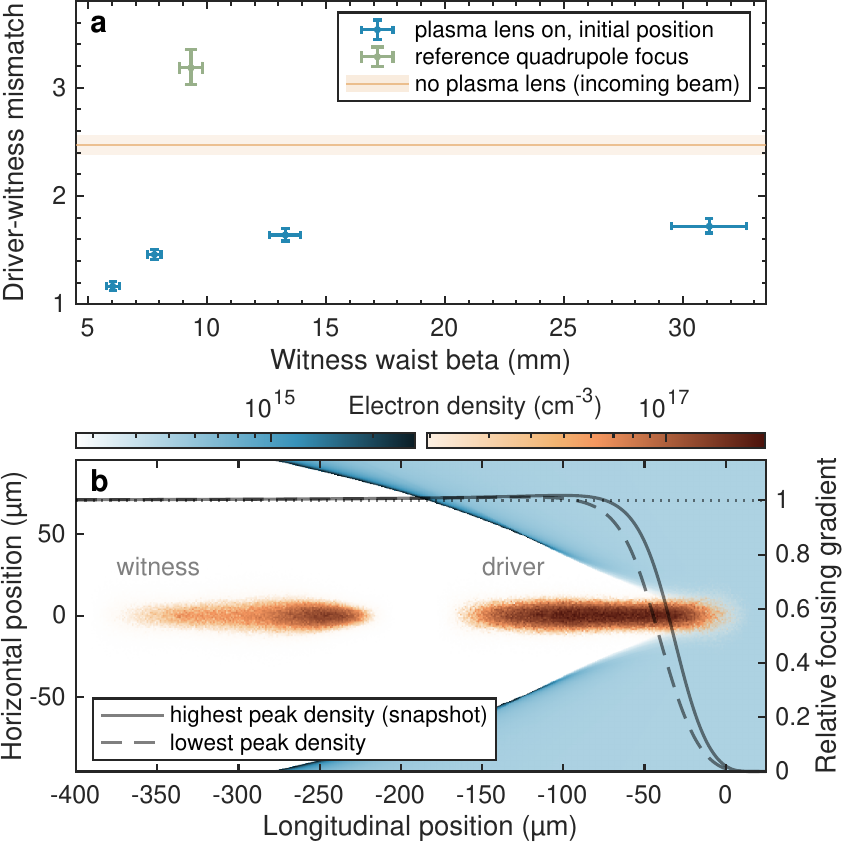}
\caption{\label{fig:6} Comparison of driver and witness optics. \textbf{a}. Measured projected mismatch parameter (Eq. \eqref{eq:mismatch}) between driver and witness as a function of the witness's waist beta. Horizontal error bars show the 95\,\% CI of the fits, vertical errors show the 95\,\% CI based on Monte-Carlo sampling of driver and witness waist parameters. \textbf{b}. Snapshot of the simulated beam-plasma interaction in the highest-density working point ($n_\mathrm{p,0}=5\times10^{14}$\,cm\textsuperscript{-3}, rightmost point in Fig. \ref{fig:5}d). Solid and dashed curves show the on-axis focusing gradient normalized to the nominal blowout gradient for the lowest and highest peak density working points. The transversely elliptical (as opposed to circular) driver causes the relative horizontal (vertical) gradients to overshoot (undershoot) 1 \cite{manwani_2025_PRL}.}
\end{figure}

\paragraph{Impact on the driver bunch}
Since the operation of a PBA requires a final-focusing PPL to focus both the witness and the driver, it is necessary to study the driver's evolution. Figure \ref{fig:6}a shows the Twiss mismatch parameter between the driver and witness bunches \cite{minty_zimmermann_2003_measurement_control,mehrling_2012_PRSTAB}, 
\begin{equation}\label{eq:mismatch}
    \mathcal{M}_\mathrm{D-W} = \frac{\beta^{*2}_\mathrm{D} + \beta^{*2}_\mathrm{W} + \left( s^*_\mathrm{D} - s^*_\mathrm{W} \right)^2 }{2\beta^{*}_\mathrm{D}\beta^{*}_\mathrm{W}},
\end{equation}
cast in waist-parameter form by assuming vacuum propagation (Eq. \eqref{eq:beamprop}) around the waists. A mismatch parameter of 1 corresponds to a perfect optics match. The smallest mismatch point is spuriously decreased by the mis-measured waist betas (rightmost data points in Fig. \ref{fig:5}b). For equal waist betas, the mismatch is significantly smaller when focusing with the PPL than with the quadrupoles. The driver would thus be more closely matched for a witness waist matched to a downstream plasma accelerator. This can reduce transverse driver evolution in the wake and, for a beam-driven plasma-wakefield accelerator (PWFA), enable a higher degree of driver charge capture, increasing acceleration stability and energy efficiency. 

Figure \ref{fig:6}b shows a snapshot from a simulation of the beam-plasma interaction at the highest-density working point ($n_\mathrm{p,0}=5\times10^{14}$\,cm\textsuperscript{-3}). While the highly overdense driver ($n_\mathrm{b,0} \approx 370\times n_\mathrm{p,0}$) led to a clear blowout, the lowest peak plasma density did not yield the most efficient wake excitation; higher plasma densities showed excitation earlier along the driver, with the highest density yielding the earliest excitation, focusing the largest fraction of the driver. The efficiency of this working point agrees with idealized simulations (see Appendix) and is interpreted as an interplay between more resonant wake excitation \cite{marocchino_2017_APL} (favoring higher plasma densities) and driver over-density (lower plasma densities). The majority of bunch charge sees nearly the nominal focusing gradient, but the varying gradient at the head results in locally varying downstream waist parameters. However, since the focusing gradient inside a PWFA varies similarly, using a PPL for final focusing might intrinsically provide favorable matching conditions for the driver bunch. The relative driver emittance growth at the highest peak-density working point was 6\,\%.

\paragraph{Discussion and outlook}
While shorter optics are theoretically less chromatic \cite{doss_2019_PRAB}, the lower-density, longer PPL regime investigated here offers advantages over the higher-density regime \cite{ng_2001_PRL,doss_2019_PRAB}. The lower plasma density reduces the bunch-density requirements on the driver bunch, allowing a large fraction of it to focus. Longer lenses also induce less synchrotron radiation aberration than shorter lenses, given identical waist parameters \cite{oide_1988_PRL,blanco_2016_PRAB}. If spatially overlapping with the adjacent PBA plasma-density ramps, long plasma lenses are perhaps more akin to independently tunable ramp extensions than separate optics, which could improve the performance of the system as a whole \cite{ariniello_2019_PRAB,chang_2023_PRA,ariniello_2022_PRR}.

The lower-density regime would also be advantageous for witness bunch capture after a PWFA, as the larger wake could allow the more divergent, energy-depleted driver tail to contribute to wake excitation. As the driver is not needed downstream of a PBA, a capturing PPL could be optimized predominantly for the witness bunch. For a PWFA-based collider, the large differences in particle energy between the driver and witness may be difficult to manage when focusing into the PWFA stages, while for FEL-type energy boosters, the driver and witness particle energies would be more similar.

In summary, we have for the first time experimentally demonstrated slice-emittance preservation in a PPL, the tunability of the waist parameters, and reduced aberration compared to focusing using quadrupole magnets. While emittance growth was observed for part of the data, the sources were identified and can be mitigated. A low-density PPL has been shown to have clear advantages as a focusing section for matching into a PWFA stage. Shown compatibility with beams three orders of magnitude brighter than previous work \cite{lindstrom_2018_PRL} marks these findings as a critical step toward compact, beam-quality-preserving optics in novel accelerators.

\begin{acknowledgments}
\paragraph{Acknowledgements}
The authors would like to thank M. Dinter, S. Karstensen, S. Kottler, K. Ludwig, F. Marutzky, A. Rahali, V. Rybnikov, A. Schleiermacher, R. Wolf, the FLASH management, and the DESY FH and M divisions for their scientific, engineering, and technical support. JBS also thanks A. Novokshonov (DESY) for helpful discussions about scintillator saturation, A. Sinn (DESY) for HiPACE++ support, and F. Curbis (Lund University, MAX IV Laboratory) for feedback on the manuscript. This work was supported by Helmholtz ARD, the Helmholtz IuVF ZT-0009 program, and the Maxwell computational resources at DESY. Grammarly was used to assist in proofreading the manuscript.
\end{acknowledgments}

\paragraph{Author contributions}
JBS and RD conceived the experiment. JBS, LB, AK, and SS performed the experiment, with help from JB, PGC, GL, FP, and SW. MH and HJ performed the plasma density measurements. JBS developed the experiment methodology, analyzed the experimental data, produced all figures, and wrote the manuscript. JBS performed the PIC simulations and analyzed the results, with help from LB and SS. RD and JW supervised the project and the personnel. All authors discussed the results in the paper.


\paragraph{Data and code availability}
The data and code that support the findings of this article are openly available \cite{bjorklund_svensson_2026_zenodo}. The codes Wake-T \cite{ferran_pousa_2019_JPhys} and HiPACE++ \cite{diederichs_2022_CompPhysCommun} are open source and freely available.

\appendix

\section{Appendices}

\paragraph{The beam emittance} A particle beam's RMS normalized emittance \cite{floettmann_2003_PRAB} is usually used to quantify the transverse beam quality, or \enquote{focusability}. The normalized emittance is defined as $\varepsilon_\mathrm{n}^2 = \langle x^2 \rangle \langle u_x^2 \rangle - \langle x u_x \rangle^2$, where $x$ denotes the distance from the nominal trajectory, and $u_x = p_x / (m_e c)$ is the transverse momentum, $p_x$, normalized by the electron mass, $m_e$, and the speed of light in vacuum, $c$. The normalized emittance is preserved under acceleration and linear focusing forces such as those in quadrupole magnets and plasma wakes in the blowout regime. A finite value implies a finite focal spot size $\sigma_x^* = \sqrt{\varepsilon_{\mathrm{n},x}\beta^*_x/\gamma_\mathrm{b}}$ (Eq. \eqref{eq:beamprop} at $s=0$) even with idealized optics.

\paragraph{Single-optic chromaticity}
The chromatic amplitude function \cite{montague_1979_LEP,lindstrom_2016_PRAB}, \mbox{$W^2 = \left( \partial_\delta\alpha - \alpha\partial_\delta\beta / \beta \right)^2 + \left( \partial_\delta\beta / \beta \right)^2$}, where $\partial_\delta = \partial/\partial\delta$, can be used to quantify the chromaticity of an ensemble of particles. It contributes to relative emittance growth to first order as $\Delta\varepsilon/\varepsilon_0 \approx W^2\sigma_\delta^2\,/\,2$ \cite{lindstrom_2016_PRAB}. A single, thin focusing optic adds a chromatic amplitude of $W\approx\beta\kappa l$ \cite{doss_2019_PRAB}, where $\beta$ is the beam's beta function at the optic. The focal length of an optic is $f=1/(\kappa l)=L^*$. The beam's beta function at the optic is given by $\beta\approx L^{*2}/\beta_\mathrm{m}$. These equations combine to yield $W \approx L^*/\beta_\mathrm{m}$ such that $\Delta\varepsilon/\varepsilon_0 \approx L^{*2}\sigma_\delta^2/(2\beta_\mathrm{m}^2)$.

\paragraph{Chromaticity correction}
Chromaticity is usually mitigated near its source by using nonlinear magnets (sextupoles) in the presence of large transverse dispersion, but nonlinear effects must be carefully managed to avoid emittance growth \cite{raimondi_2001_PRL}. However, chromaticity can also be corrected by non-dispersive \enquote{apochromatic} optics, which scale favorably in terms of both energy acceptance and length when using strong plasma lenses \cite{lindstrom_2016_PRAB}. Because PPLs are less chromatic, less correction is needed.

\paragraph{Imaging spectrometer}
The high-resolution imaging spectrometer was made up of five high-gradient quadrupoles with an 8-mm bore radius, a 1.07 m long vertically dispersive dipole magnet, and a European XFEL-type screen station \cite{kube_2015_IBIC} with an in-vacuum GAGG:Ce scintillating screen \cite{kube_2019_FEL}. The image resolution, limited by the point spread function (PSF) of the imaging system, was found to be 6.7\,$\mu$m RMS with the same method as in Ref. \cite{lindstrom_2024_NatCommun}. A charge calibration was obtained by scanning one of the energy collimators and correlating the image pixel count with an upstream toroidal current transformer. Local scintillator saturation was compensated by applying Birks' Law \cite{Birks_Scintillators,kurz_2018_RSI}; the actual electron-beam charge density, $\rho_\mathrm{b}$, was retrieved through $\rho_\mathrm{b} = \rho_\mathrm{s}/(1-B\rho_\mathrm{s})$, where $\rho_\mathrm{s}$ is the charge density obtained directly from the scintillator and $B$ is Birks' constant for the material. A value of $B=0.08$\,mm\textsuperscript{2}/nC was found, corresponding to a saturation charge density of 12.5\,nC/mm\textsuperscript{2}. Critically, because of the small slice beam sizes encountered in this study, the images were deconvolved with the 2D PSF before applying the saturation compensation, as that correction assumes that $\rho_\mathrm{s}$ is given by the light yield from the scintillator, which differs from the camera image when resolution-limited. Therefore, the resolution term used in Ref. \cite{lindstrom_2024_NatCommun} was omitted when fitting to the object-plane scans. In the dispersive plane, the PSF was a 2nd-order super-Gaussian with a 34.7\,$\mu$m RMS because of the angle of 45 degrees between the optical axis and the scintillator \cite{kube_2015_IBIC}. The RMS energy resolution was estimated as $\sigma_\delta\gg\sigma_{y_0}/D_y$, where $\sigma_{y_0}$ is the undispersed beam size and $D_y=0.15$\,m is the linear dispersion. Assuming similar waist sizes to $x$ (c.f. Fig. \ref{fig:4}a,d), this yields $\sigma_\delta\gg2\times10^{-5}$, which is fulfilled with 0.25-MeV slices ($\Delta\delta_{\mathrm{slice}}\approx2.4\times10^{-4}$). Conversely, a 0.25-MeV slice requires an undispersed beam size of $\sigma_{y_0}\ll36\,\mu$m, which is usually fulfilled even if not exactly imaging the slice waist due to slice Twiss variation.

\paragraph{Beam reconstruction}
The 4D ($x,x^\prime,z,\delta$) beam phase space before beam-plasma interaction was reconstructed using energy slice parameters from the high-resolution spectrometer in combination with an LPS measured using the X-band transverse deflecting structure \cite{gonzalez_caminal_2024_PRAB}. The monotonic LPS shape allowed a direct mapping between energy and time for the horizontal plane. For the vertical plane, only the projected beam parameter could be measured, and thus the ($y,y^\prime$) slice values were approximated from these measurements. Figure \ref{fig:2} shows the reconstructed LPSs and horizontal slice parameters.

\paragraph{PIC simulations} All PIC simulations were performed with HiPACE++ following preliminary studies with Wake-T \cite{diederichs_2022_CompPhysCommun,ferran_pousa_2019_JPhys}. Simulation details are listed according to the figure in which the results are shown.

\noindent\textbf{Figure \ref{fig:1}} The bunches in these idealized simulations were transversely symmetrized to suppress numerical aberrations. The simulation parameters were otherwise scaled as for Fig. \ref{fig:4}, see below. The starting waist beta was 10 mm. Gaussian bunches with FLASHForward-like parameters (see Fig. \ref{fig:2}) were used, apart from the driver emittance, which was set to 8\,mm$\cdot$mrad to highlight the lens operating limits.

\noindent\textbf{Figure \ref{fig:4}} These simulations were performed with beams reconstructed from experimental data, where the beam shown in Fig. \ref{fig:2}a was used for the Fig. \ref{fig:4} simulations. The vertical slice parameters were set to the projected values, apart from the emittance which was halved to decrease the driver ellipticity causing large longitudinal variations in the focusing gradient (similar to Fig. \ref{fig:6}b, but significantly larger). Both bunches contained $10^7$ macroparticles (MPs). The simulation box had $4095\times4095\times525$ grid cells $(x,y,z)$, with 2 MPs per cell within a variable (with peak plasma density) radius around the wake (minimum radius $10\times\sigma_{r,\mathrm{driver}}$), and 1 MP per cell outside. The transverse box size was set to $2/k_\mathrm{p}$ (on the mm scale) as this was observed to be large enough not to affect the fields. The transverse resolution was thus variable, but with a minimum $\sigma_{r,\mathrm{witness}}/10$. The longitudinal resolution was 1 $\mu$m. The simulation step length was adaptive with a maximum of $\min(\, l/10, \,\lambda_\mathrm{p}/120 \,)$. The plasma distribution was nearly identical to that in the numerical model in Fig. \ref{fig:4}; the density was reduced by 7.5\,\% to match the experiment because of the focusing gradient overshoot from the elliptical driver. Ion motion was verified not to impact the results.

\noindent\textbf{Figures \ref{fig:5} and \ref{fig:6}b} Mostly identical setup to that for Fig. \ref{fig:4}, but with a plasma of RMS length 7.5\,mm to account for argon being heavier than nitrogen and being expelled more slowly out of the capillary. The beam in Fig. \ref{fig:2}b was used. The peak plasma densities were logarithmically spaced on the span $5\times10^{13}$\,--\,$5\times10^{14}$\,cm\textsuperscript{-3}, corresponding to focusing gradients of approximately 1.5\,--\,15\,kT/m. The real RMS length of the plasma was expected to change somewhat over time, but this effect was omitted for simplicity. The timings of simulated results relative to the experimental data were manually fitted. Simulations were also performed with a finite plasma temperature (1\,eV) and beam-plasma and plasma-plasma collisions enabled to investigate any impact on the driver. Only minor effects were observed, which we attribute to the modest temperature and low electron/ion density. 

\begin{figure}[b!]
\includegraphics[width=1.0\linewidth]{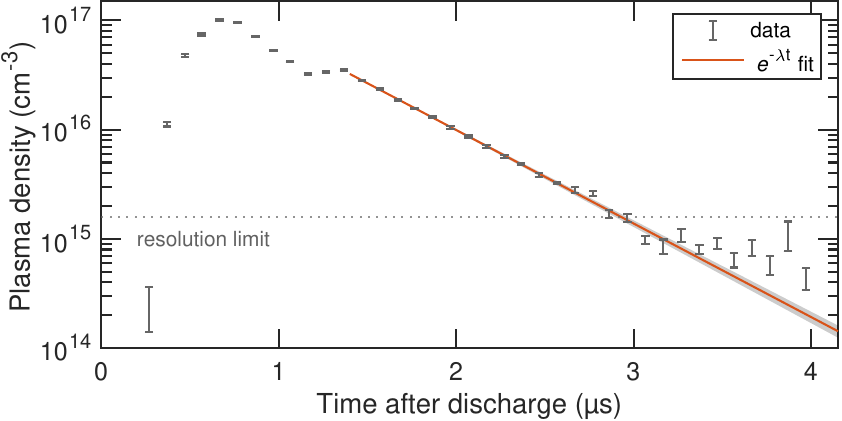}
\caption{\label{fig:A_1} Plasma density data from OES measurements. }
\end{figure}

\paragraph{Plasma density measurements} Plasma density measurements were performed offline using optical emission spectroscopy (OES) \cite{garland_2021_RevSciInstr}. The results are shown in Fig. \ref{fig:A_1}, where it is evident that the plasma densities of interest (c.f. Fig. \ref{fig:1}a) are below the resolution limit of $1.6\times10^{15}$\,cm\textsuperscript{-3}, preventing direct use of OES data for determining the PPL plasma characteristics. The plasma density profile used in Fig. \ref{fig:4} was obtained from the best fit to the focused beam parameters and not from OES measurements.

\bibliography{a_references}

\end{document}